\documentclass[aps,prl,twocolumn,superscriptaddress,showpacs,floatfix]{revtex4-1}
\usepackage{amssymb}
\usepackage{amsmath}
\usepackage{bm}
\usepackage{url}
\usepackage{color}
\usepackage{graphicx}

\begin{document}
\title{Defects, parcellation, and renormalized negative diffusivities in non-homogeneous oscillatory media}

\author{Marie Sellier-Prono}
\affiliation{Laboratoire de Physique, Ecole Normale Sup\'erieure, CNRS, PSL Research University, Sorbonne University, Paris 75005, France}
\author{Massimo Cencini}
\affiliation{Istituto dei Sistemi Complessi, CNR, via dei Taurini 19, 00185 Rome, Italy
and  INFN, sez. Roma2 ``Tor  Vergata''}
\author{David Kleinfeld}
\affiliation{Department of Physics, University of California San Diego, La Jolla, CA 92093, USA}
\affiliation{Department of Neurobiology, University of California San Diego, La Jolla, CA 92093, USA}
\author{Massimo Vergassola}
\affiliation{Laboratoire de Physique, Ecole Normale Sup\'erieure, CNRS, PSL Research University, Sorbonne University, Paris 75005, France}
\affiliation{Department of Physics, University of California San Diego, La Jolla, CA 92093, USA}

\begin{abstract}
Spatial non-homogeneities can synchronize clusters of spatially-extended oscillators in different frequency plateaus. Motivated by physiological rhythms, we fully characterize the phase diagram of a Ginzburg-Landau (GL) model with a gradient of frequencies. For large gradients and diffusion, the rest state is stable, and the linear spectrum around it maps onto the non-Hermitian Bloch-Torrey equation. When complex pairs of eigenvalues turn unstable, precursors of plateaus grow, separated by defects where the GL amplitude vanishes. Nonlinear effects either saturate the amplitude of plateaus or lead to a phase-locked state, with saddle-node bifurcations separating the two regimes. In the region of plateaus, we trace the formation of defects to a non-linear renormalization of the diffusivity, and determine the scaling of their number and length {\it vs} dynamical parameters.
\end{abstract}


\maketitle 

Spatially-extended oscillations appear in a variety of situations, ranging from engineering to living, physical, and social systems \cite{Kuramoto1984,Pikovsky2001,Strogatz2004}. Hallmarks of nonlinearities and spatial couplings that characterize those systems are synchronization among different oscillators, and entrainment by external forcing.  Synchronization is defined as the oscillation in unison of oscillators that have different frequencies yet lock because of their mutual coupling. A spectacular example is the synchronized flashing of thousands of firefly males during mating season \cite{Buck1966,Peleg2020}. Other well-studied examples include locomotion, especially by fish~\cite{Grillner2020}, and electrical waves in neural computations~\cite{Ermentrout2001, Muller2018}. Entrainment by external forcing is the locking of oscillators to the driving rhythm. A physiological instance that impacts functional magnetic resonant imaging (fMRI) is vasomotion~\cite{Haddock2005,Drew2020}. Arterioles that source blood to the brain have the diameter of their walls driven by oscillations of the surrounding smooth muscle cells. In the presence of external neuronal stimuli, a strong vasomotion increase is observed at the drive frequency~\cite{Broggini2024}.

A number of theoretical methods have been developed to grasp the above phenomena. Phase models~\cite{Kuramoto1984,Pikovsky2001} are rooted in the insight that weak couplings among oscillators perturb the stable limit cycle of each oscillator in a specific manner: Transverse perturbations relax rapidly, while phase changes along the cycle can be large and do not relax (zero modes). Technically,
the limit cycle and its isochrones \cite{Winfree2001} are unperturbed, while weak perturbations  drive the phase. The major simplification is that each oscillator is described by a single variable, its phase in the limit cycle, rather than the original, generally multidimensional, description (see Supplementary Material (SM)~\cite{NoteSM}). The famous Kuramoto transition to synchronization \cite{Kuramoto1984} was obtained within this framework.

\begin{figure}[t!]
\centering
\includegraphics[width=1\columnwidth]{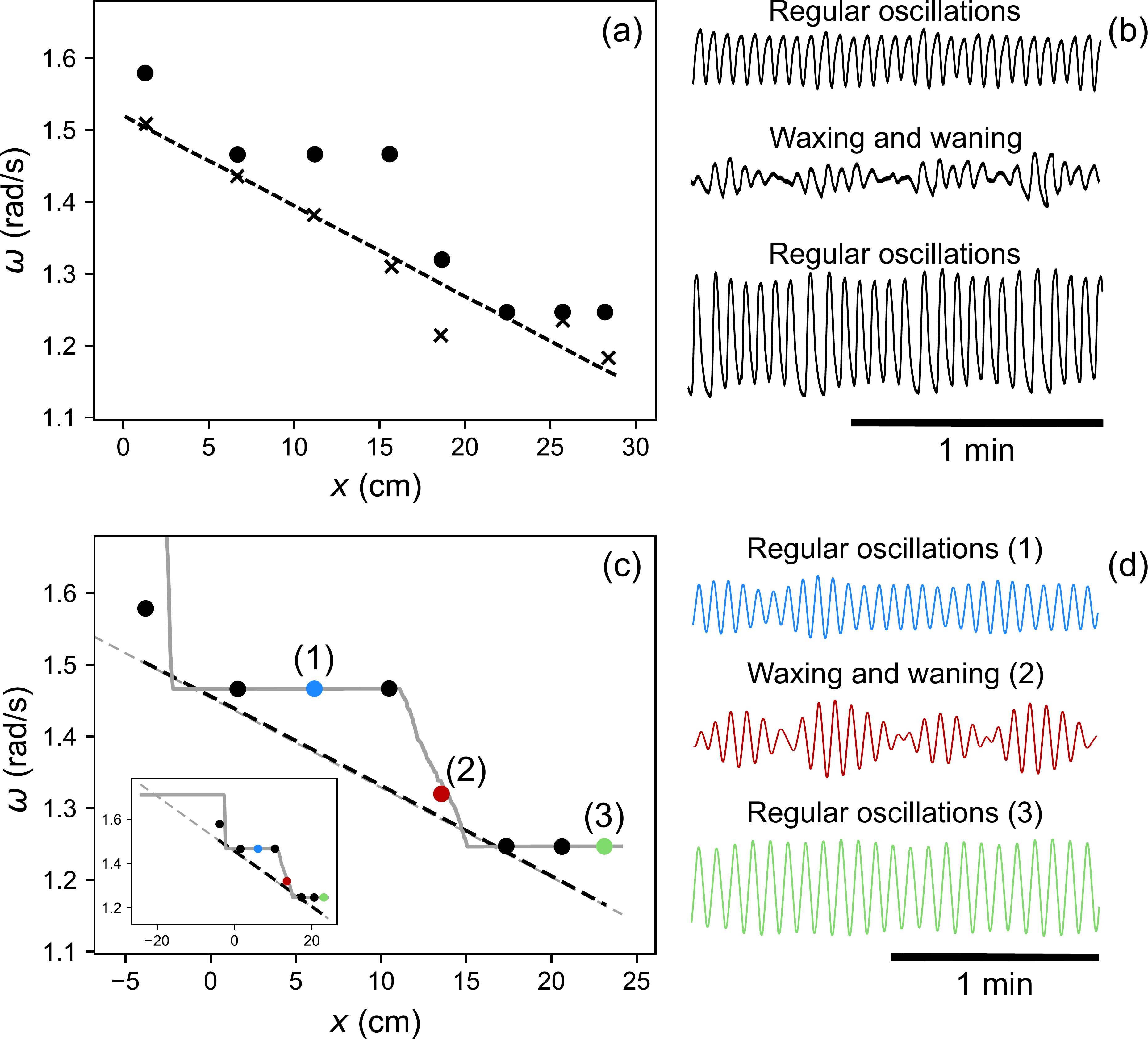}
\vspace{-0.5truecm}
\caption{(Color online) Experimental peristalsis and comparison of our GL modeling. (a) Average frequencies recorded in isolated  segments of cat intestine (crosses\,; the black dashed line is their linear fit) and in the intact intestine (circles); (b) Time-course  signals within two distinct plateaus (top and bottom), where relatively regular oscillations are observed, and in the separating region of ``waxing and waning''. Data were extracted from Ref.~\cite{Diamant1969} as detailed in \cite{NoteSM}. Panels (a) and (b) refer to two different experimental realizations so we do not associate the curves in (b) to specific points in (a). Panels (c) and (d) show the same curves as in (a) and (b) obtained from our GL modeling with the use of the procedure detailed in \cite{NoteSM}. The plateau architecture is captured, as well as the structure of the three time courses that we locate by their colors. The inset shows the full simulation domain.}
\label{fig1}
\end{figure}

We focus on spatially non-homogeneous oscillatory media. Non-homogeneity may be caused by spatial variations of the parameters, e.g., the frequency of oscillators, or non-uniform external drive. Our motivation stems from two natural systems under intense experimental investigation\,: the aforementioned cortical vasomotion~\cite{Haddock2005,Drew2020,Broggini2024}, and peristalsis~\cite{Diamant1969,Parsons2020}. Interstitial cells in the small intestine of mammals  bear spontaneous oscillations that are synchronized by gap junction couplings, and generate rhythmic waves of contraction. Non-homogeneity is imposed by the frequencies of the oscillators in isolation decreasing aborally in a roughly linear fashion from the stomach to the colon (see Fig.~\ref{fig1}a), as found by {\it post hoc} sectioning of the intestine \cite{Diamant1969}.  In the intact intestine, couplings result in parcellation, i.e., a series of  frequency plateaus as in Fig.~\ref{fig1}a.  Oscillations within the plateaus are regular, while their boundary displays ``waxing and waning'' (see Fig.~\ref{fig1}b) that results from the interference of mismatched adjacent cell populations~\cite{Diamant1969}. 

To understand parcellation, it is natural to consider a chain of weakly coupled oscillators with a gradient of natural frequencies. Ref.~\cite{Ermentrout1984} pioneered this approach and demonstrated the existence of plateaus. However, as already noted in \cite{Ermentrout1986}, the strong amplitude modulation at the boundary of plateaus is incompatible with the weak coupling approximation. A first alternative is to consider chains of nonlinear oscillators, {\it viz.}, van der Pol type \cite{Linkens1974,Osipov1998,Menzel2011}, which is also relevant for density waves in dusty plasmas \cite{Menzel2011bis}. Numerical simulations confirmed the existence of plateaus and evidenced  ``oscillator death'' for large enough gradients, i.e., the presence of regions with oscillators that are unexcited \cite{BarEli1985,Ermentrout1990,Reddy2000,Ozden2004}. A second alternative is to consider a Ginzburg-Landau (GL) model (see \cite{Aranson2002}) with a gradient of natural frequencies. The advantage is that the GL equation is the universal normal form close to a supercritical Hopf bifurcation. Oscillations are described by a complex field with a phase and an amplitude, in contrast to phase models. Refs. \cite{Ermentrout1986,Ermentrout1989,Akopov2003,Anishchenko2005} numerically demonstrated plateaus in the GL equation and analyzed its locked states
and amplitude death at large gradients (see below). In the GL formulation, the ``waxing and waning'' at the boundary of plateaus is associated with defects, i.e., space-time locations where the complex field vanishes and its phase is undefined \cite{Pikovsky2001}. 

\begin{figure*}[bht]
\centering
\includegraphics[width=0.8\textwidth]{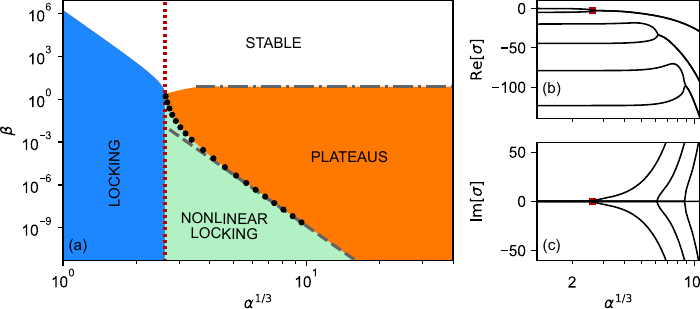}
\vspace{-0.3truecm}
\caption{Phase space portrait (a) and eigenvalues of the linear spectrum around the rest state $A=0$ (b-c). (a) Stable: The rest state is stable (and unique) in the white region at high $D$ and $g$ \cite{Ermentrout1989}. All eigenvalues have negative real part. The dashed-dotted line is the instability curve at large $\alpha$, which we predict using results for the spectrum in a semi-infinite domain \cite{Dyson1991}. 
Plateaus: at least a complex pair of eigenvalues has a positive real part (linear precursors of plateaus) and the nonlinear regime features multiple plateaus. Locking and nonlinear locking: A locked state \cite{Ermentrout1986} is observed in the nonlinear regime for both regions and at least one eigenvalue of the linearized spectrum has positive real part. Their difference is that the blue region has all eigenvalues real, while the green region has at least one pair of complex eigenvalues. In other words, the blue/green regions are respectively on the left/right of the branching point shown in red in panels (b) and (c) and represented by the red dotted line in panel (a). That differentiates the approach to the locked state between the two regions (see Fig.~S9). 
The black dashed line is the asymptotics for the fold bifurcation curve separating nonlinear locking from plateaus, as we predict by Eq.~\eqref{eq:critical}. (b)-(c): the real and imaginary part of the first six eigenvalues $\sigma_n$ {\it vs} $\alpha^{1/3}$ for the Bloch-Torrey operator defined by Eq.~\eqref{eq:BlochTorrey} (ordered by their real part). The imaginary part vanishes before the branching points (the first one is in red) while two complex conjugate eigenvalues per branch are present afterwards.}
\label{fig2}
\end{figure*}

We revisit the GL equation with a gradient of natural frequencies to characterize its phase diagram (see Fig.~\ref{fig2}). 
Specifically, we consider the one-dimensional Ginzburg-Landau (GL) equation for the complex field $A(x,t)$\,: 
\begin{equation}
    \label{eq:GL}
    \partial_t A(x,t) = rA(1 - |A|^2) + igxA + D \Delta A\,,
\end{equation}
which is the Stuart-Landau normal form for oscillators close to a supercritical Hopf bifucation, with diffusion accounting for spatial local coupling terms \cite{Kuramoto1984,Pikovsky2001}. We rescaled the amplitude of $A$ to match the coefficients of the cubic and the instability term, $r$, which we take real and positive, along with the diffusivity $D$. No-flux boundary conditions entail $\nabla A\left(\pm L/2\right)=0$, the size of the domain being $L$. Note that the cubic nonlinearity is dictated by gauge invariance $A\mapsto Ae^{i\phi}$. The linear gradient $gx$ imposes non-homogeneity and breaks parity. That allows {\it a priori} for an additional term proportional to $\nabla A$ that we shall consider later, together with complex $r$ and $D$, to capture experimental data in Fig.~\ref{fig1}.

We start by the simplest possible solution of Eq.~\eqref{eq:GL}, i.e., the rest state $A=0$, which is known to be stable and unique at high values of $D$ and $g$ \cite{Ermentrout1989}. The physics is that $g$ tends to create phase differences among neighboring oscillators, while $D$ tends to homogenize their state. For strongly-driven phase differences, the only possibility left to close-up the gap in states is to shrink the limit cycle, and eventually drive it to zero \cite{BarEli1985}. To identify the term in Eq.~\eqref{eq:GL}  responsible for reducing the amplitude as phase differences increase, we decompose the field as $A(x,t)=a(x,t)e^{i\varphi(x,t)}$ and rewrite Eq.~\eqref{eq:GL} as\,:
\begin{eqnarray}
    \label{eq:GL_a}
    \partial_t a(x,t) &=&  r a (1 - a^2) - D a (\nabla\varphi)^2 + D\Delta a\,,
    \\
    \partial_t \varphi(x,t) &=& gx + 2 D \left(\nabla \ln a\right) \left(\nabla \varphi\right) + D \Delta \varphi\,,
    \label{eq:GL_phi}
\end{eqnarray}
which makes explicit the damping term $- D a (\nabla\varphi)^2$.

In the linear regime around the rest state, $|A|\ll 1$, Eq.~\eqref{eq:GL} reduces to
\begin{equation}
    \partial_\tau A = \partial^2_\zeta A + (r^\prime + i\alpha\zeta) A\,;\quad r^\prime\equiv \frac{rL^2}{D},\,\,\, \alpha\equiv \frac{|g|L^3}{D}\,,
    \label{eq:BlochTorrey}
\end{equation}
where $\zeta\equiv x/L$ and $\tau\equiv Dt/L^2$. Ref.~\cite{Ermentrout1989} showed that the eigenvalues of Eq.~\eqref{eq:BlochTorrey} 
must have a negative real part if $\alpha$ is large and $r'$ is small enough because the alternative hypothesis leads to a contradiction.  An alternative path is to 
recognize that Eq.~\eqref{eq:BlochTorrey} for $r'=0$ was introduced by Bloch \cite{Bloch1946} and Torrey \cite{Torrey1956} for nuclear magnetic resonance and its spectrum was characterized in Ref.~\cite{Dyson1991}. The eigenvalues $\lambda_n$ of Eq.~\eqref{eq:BlochTorrey} are the Bloch-Torrey eigenvalues $\sigma_n$'s shifted as  $\lambda_n = \sigma_n + r^\prime$ (see End Matter). Since the $\sigma$'s have a negative real part (see Fig.~\ref{fig2}), the $\lambda_n$'s 
remain negative for $r^\prime < {\rm min}(-\mathrm{Re}[\sigma_n])$, in agreement with Ref.~\cite{Ermentrout1989} and the phase diagram of Fig.~\ref{fig2} as replotted in the variables $(\alpha,r')$ in the supplementary Fig.~S7~\cite{NoteSM}.

The above alternative path brings two more results. First, 
it yields an analytical prediction for the instability curve at large $\alpha$'s. The difficulty with imposing the two no-flux boundary conditions in Eq.~\eqref{eq:BlochTorrey} is that its eigenfunctions are Airy functions \cite{ValleeAiry2004,Abramowitz1965}. In the limit of large $\alpha$'s, though, one can treat the domain as semi-infinite and use properties of zeros of Airy functions \cite{NoteSM,Dyson1991}. The only dimensional combination of $g$ and $D$ (without $L$) to form an inverse time is $\left(g^{2}D\right)^{1/3}$. The upshot (see \cite{NoteSM}) for our Eq.~\eqref{eq:BlochTorrey} is that we predict the rest state should become unstable for $r-0.5094\left(g^{2}D\right)^{1/3}>0$, i.e., the dash-dotted line in Fig.~\ref{fig2}\,:
\begin{equation}
\label{eq:dashdottedline}
\beta\equiv \frac{Dg^2}{r^3}< \beta_c\simeq \left(\frac{1}{0.5094}\right)^3\simeq 7.565\,.
\end{equation}
Second, the alternative path identifies branch points of the non-Hermitian operator in Eq.~\eqref{eq:BlochTorrey}. Branch points are defined as points where two eigenmodes fuse~\cite{Grebenkov2022}, as in Figs.~\ref{fig2}b,c for the Bloch-Torrey operator~\cite{Dyson1991}. For $\alpha=0$, the eigenvalues are real (the operator is Hermitian). As $\alpha$ increases, pairs of branches interconnect for successive values $\alpha_p$ and become complex conjugate for  $\alpha > \alpha_p$, with the first branching at $\alpha_1\approx18.1$~\cite{Dyson1991}. We show in the End Matter that conjugate pairs are associated with precursors of defects, i.e., strong amplitude modulation with a period $2\pi/|\mathrm{Im}\, \lambda_1|$. The first branching value $\alpha_1^{1/3}\simeq 2.62$
identifies the cusp at the intersection among locking, plateaus and rest states in Fig.~\ref{fig2}. 

\begin{figure*}[bht!]
\centering
\includegraphics[width=1\textwidth]{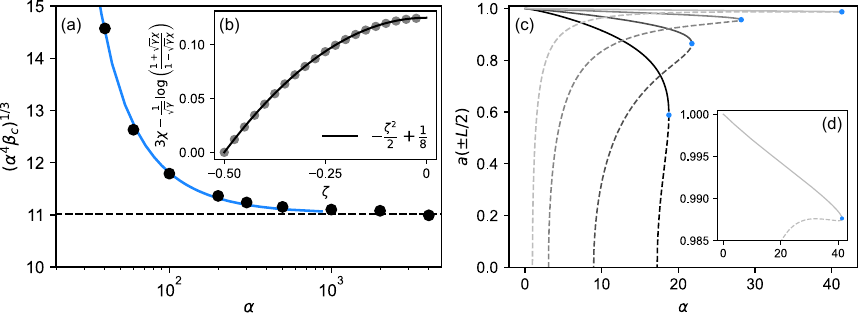}
\vspace{-0.5truecm}
\caption{Nonlinear phase-locking and its transition to
plateaus.
(a) The critical curve $\alpha^{4/3}\beta_c^{1/3}$ \textit{vs} $\alpha$
as identified by numerical integration of the GL equation (black points) and
by the location of the saddle-node bifurcation (blue line).
Inset (b): The l.h.s. of Eq.~\eqref{eq:solve} \textit{vs} the rescaled space
variable $\zeta$ (grey circles) obtained from simulations for
$\alpha=2000$, $\beta=8.50\times 10^{-11}$ and the predicted r.h.s. of Eq.~\eqref{eq:solve} (solid line).
(c) Bifurcation diagram of the solutions to Eqs.~\eqref{eq:GL_aI} and \eqref{eq:GL_phiI} specialized to the locked case (see text) for $\beta=1, 10^{-1},
10^{-2}, 10^{-3}$ (from darkest to lightest). Points in blue are the bifurcation points, as for the line in (a).
Inset (d): zoom of the uppermost $\beta=10^{-3}$ curve.}
\label{fig3}
\end{figure*}

In the unstable region (${\rm Re}\,\, \lambda_1>0$), exponential amplification occurs and nonlinear effects become relevant. On the left of the cusp, eigenvalues are all real 
and diffusion is strong enough to couple the whole interval, leading to the phase-locked state unveiled in Ref.~\cite{Ermentrout1986}. The amplitude of the locked state is progressively reduced 
as the boundary with the stable region is approached, where it eventually vanishes~\cite{Ermentrout1986}. As $\alpha$ increases, the range of diffusion shrinks, and the locked state breaks up into multiple plateaus~\cite{Ermentrout1986}. To 
determine the nature of the change in regime, we note that Fig.~\ref{fig3}c highlights the necessity of nonlinear effects as the transition does not occur at $\alpha_1$, as it does in the linear regime. Specifically,  inspection of the critical curve suggests its scaling as $\beta_c(\alpha)\propto \alpha^{-4}$, which motivates the definition
\begin{equation}
\gamma\equiv \left(\beta\alpha^4\right)^{\frac{1}{3}}=\frac{L^4g^2}{rD}\,.
\label{eq:gamma}
\end{equation}
Locking occurs for $\gamma<\gamma_c$ and the critical values asymptote to $\gamma_c\simeq 11.02$ at large values of $\alpha$ (see Fig.~\ref{fig3}a).

The above hints are captured and used to understand the transition as follows. Equations~\eqref{eq:GL_a} and \eqref{eq:GL_phi} are conveniently non-dimensionalized as
\begin{eqnarray}
    \label{eq:GL_aI}
    \partial_{\tau} a &=&  \alpha^2 a\left[ \frac{1 - a^2}{\gamma} - (\varphi^{\prime})^2\right] + a^{\prime\prime}\,,
    \\
    \partial_{\tau} \varphi &=& \zeta + 2 \left(\log a\right)^{\prime} \varphi^{\prime} + \varphi^{\prime\prime}\,, 
    \label{eq:GL_phiI}
\end{eqnarray}  
where space and time are rescaled as in Eq.~\eqref{eq:BlochTorrey}, and primes denote $\zeta$-derivatives. For large values of $\alpha$, $a\simeq \sqrt{1-\gamma\varphi'^2}$ and, inserting this into Eq.~\eqref{eq:GL_phiI}, we derive that nonlinearities renormalize the diffusivity
\begin{equation}
        \partial_{\tau} \varphi = \zeta + \varphi^{\prime\prime}\,\frac{1-3\gamma\varphi'^2}{1-\gamma\varphi'^2}\,.
    \label{eq:negative}
\end{equation} 
For $3\gamma\varphi'^2>1$, the diffusivity turns negative and an instability is triggered at $\zeta=0$, where the maximum value of $\varphi^{\prime}$ is located. Indeed, $\varphi^{\prime}(-1/2)=0$ (no flux) and, if the system is stable, Eq.~\eqref{eq:negative} implies $\varphi^{\prime\prime}\propto -\zeta$. The instability threshold is calculated in the stationary limit of
Eq.~\eqref{eq:negative}, i.e., $\chi^{\prime}\,\frac{1-3\gamma\chi^2}{1-\gamma\chi^2} =-\zeta$ for $\varphi^{\prime}\equiv \chi$. Integrating between $-1/2$ and $\zeta$, we obtain 
\begin{equation}
3\chi-\frac{1}{\sqrt{\gamma}}\log\left(\frac{1+\sqrt{\gamma}\chi}{1-\sqrt{\gamma}\chi}\right)=-\frac{\zeta^2}{2}+\frac{1}{8}\,. 
            \label{eq:solve}
\end{equation} 
The critical threshold $\gamma_c$ is finally derived by imposing $3\gamma_c\chi(0)^2=1$, which yields 
\begin{equation}
\gamma_c=\left[8\left(\sqrt{3}-\log\left(\frac{\sqrt{3}+1}{\sqrt{3}-1}\right)\right)\right]^2\simeq 11.027\,, 
            \label{eq:critical}
\end{equation} 
in agreement with numerics (see Figs.~\ref{fig3}a-b). Once the threshold is passed, $\varphi^{\prime}$ at the origin grows rapidly, which reduces the amplitude as $a=\sqrt{1-\gamma\varphi'^2}$. The final stages of the process involve time-dependent effects\,; it is still interesting to note that the renormalized diffusivity in Eq.~\eqref{eq:negative} diverges when $a=\sqrt{1-\gamma\varphi'^2}=0$, i.e., at the formation of the defect. 

Knowledge of the critical curve allows us to determine the type of transitions taking place along it. Equations~\eqref{eq:GL_aI} and \eqref{eq:GL_phiI} for the locked case have their left-hand sides vanishing (see \cite{NoteSM}). The amplitude is indeed stationary and even in space, and the phases increase with a common frequency $\omega$, which vanishes as $\varphi(x)$ is odd. The no-flux boundary
conditions leave unknown the boundary amplitudes that satisfy the constraint $a(-L/2)=a(L/2)$. Solutions, i.e., the values of $a(-L/2)$ reported in Fig.~\ref{fig3}c-d, are sought by a shooting method and the software XPPAUT~\cite{xppaut2003}, which allows to follow solutions as parameters vary. Figure~\ref{fig3} summarizes the conclusions\,: 
two locked solutions are present, one stable (selected by the dynamics) and one unstable, that merge at the boundary between nonlinear phase locking and plateaus. 
The presence of two solutions and their merging indicates the saddle-node nature of the bifurcations along the critical curve.   

\begin{figure}[t!]
\centering
\includegraphics[width=.9\columnwidth]{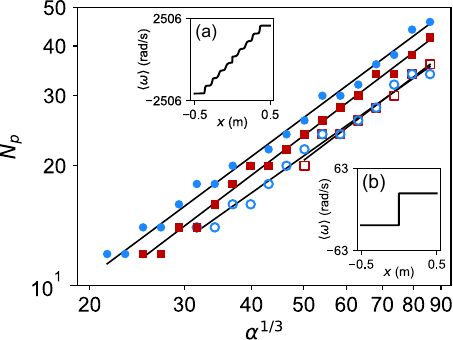}
\vspace{-0.3truecm}
\caption{(Color online) Properties of plateaus. The number of plateaus $N_p$ is shown {\it vs} $\alpha^{1/3}$ for Dirichlet (circles) and no-flux (squares) boundary conditions. Empty/full symbols refer to $\beta=10^{-2}$ and $\beta=10^{-4}$, respectively. The insets show plateaus for no-flux boundary conditions with $L=1$, $\beta=10^{-4}$ and (a) $\alpha=1002$ or (b) $\alpha=252$.}
\label{fig4}
\end{figure}

In the last region of the phase diagram (orange in Fig.~\ref{fig2}), at least a pair of eigenmodes grows linearly and stabilizes by nonlinear effects into multiple plateaus (see Fig.~\ref{fig4}). The simplest geometry has two sharp plateaus, as in Fig.~\ref{fig4}c. Each plateau features an average frequency, i.e., $\omega_1$ and $\omega_2$, respectively, common to all its oscillators. That ensures that their phases vary in time with the same dominant term yet may not keep a fixed phase difference as in strict phase locking \cite{Pikovsky2001}.
The jump $\Delta \omega=|\omega_1-\omega_2|$ dictates a characteristic time $2\pi/\Delta \omega$ for the two plateaus to accumulate phase differences. As phase gradients increase at the boundary of the plateaus, the amplitude damping term in Eq.~\eqref{eq:GL_a} drives down the amplitude.  When the difference of the phases in the two plateaus reaches $2\pi$, a defect $A=0$ is produced, the phase is undefined, and the $2\pi$ difference is reset to restart the periodic process (see Movie~1 in Ref.~\cite{NoteSM}). As already noted, increasing $\alpha$ reduces the range of diffusion and increases the number of plateaus $N_p$, as shown in Fig.~\ref{fig4}. 
Numerical results identify a clean scaling $N_p \propto \frac{L}{\ell_p}=\alpha^{1/3}$, which is independent of boundary conditions as it is also observed for 
Dirichlet boundary conditions $A\left(\mp L/2\right)=e^{\mp i gL/2}$. To  understand the $1/3$ scaling, we observe that  uncoupled oscillators spaced by 
$\ell$ accumulate phase differences at a rate $|g|\ell$, reaching differences $\simeq 2\pi$ over times $\tau_{\ell}\propto (|g|\ell)^{-1}$. Diffusive couplings extend their 
spatial range as the square-root of time. For two points to be coupled strongly enough to lock, it is necessary that $\ell\propto \sqrt{D\tau_{\ell}}$. We 
conclude that the typical length of the plateaus is $\ell_p\equiv \left(\frac{D}{|g|}\right)^{1/3}$ and
 \begin{equation}
 N_p \propto \frac{L}{\ell_p}=\alpha^{1/3}\,.
 \label{eq:scalingN}
 \end{equation} 
 The scaling~\eqref{eq:scalingN} is consistent with each point oscillating independently ($\ell_p\to 0$) in the uncoupled $D\to 0$ limit.
 The range of validity of the scaling~\eqref{eq:scalingN} depends on $\beta$, defined in Eq.~\eqref{eq:dashdottedline}.
 As $\beta$ increases for a fixed large $\alpha$  (or $\alpha$ decreases), $N_p$ is reduced, as confirmed in a limited range in Fig.~\ref{fig4}, and globally by the $(\alpha,\beta)$ supplementary plots of Fig.~S5~\cite{NoteSM}, until the scaling~\eqref{eq:scalingN} breaks down.  

Our last step is to challenge the GL model to reproduce peristalsis data in Fig.~\ref{fig1}. While ``waxing and waning'' and parcellation are easily captured, having the plateaus above the frequencies of the uncoupled system at the corresponding spatial locations is not captured by Eq.~\eqref{eq:GL}. However, as we noted, Eq.~\eqref{eq:GL} with real coefficients is not the most general form consistent with the presence of a gradient. First, since parity is broken by $g$, an additional 
term $V\nabla A$ is possible by symmetry. Such term represents advection to the right/left depending on the sign of $V$, which may occur naturally in the physiological examples mentioned in the introduction. Indeed, peristalsis~\cite{Diamant1969,Parsons2020} is  associated with aboral flows, and vasomotion with the cortical flow of blood~\cite{Haddock2005,Drew2020,Broggini2024}. Advection displaces the plateaus, which may thus be raised/lowered with respect to their Eulerian counterparts. Second, both $r$ and $D$ may be complex. Complex $r$'s are known to shift the frequency of oscillation of the Stuart-Landau equation \cite{Pikovsky2001} but they are not associated to the coupling among oscillators. In other words, they cannot account for the differences observed in Fig.~\ref{fig1} between the intact and the {\it post hoc} sectioning of the intestine \cite{Diamant1969}. 
Conversely, complex $D$'s do involve coupling among different oscillators and can raise/lower the frequency of the plateaus. Allowing for a combination of complex $D$ and advection leads to the graphs in Fig.~\ref{fig1}, which capture the main experimental features. Since the same holds for graphs obtained individually with advection or complex $D$~\cite{NoteSM}, their disentanglement calls for additional experiments.    

In conclusion, we characterized the dynamics that leads to parcellation in non-homogeneous oscillatory systems.  While our analysis specifically applies to a gradient of natural frequencies, as in peristalsis, we expect that similar considerations apply to forced cases~\cite{Diamant1969,Parsons2020}; see Ref.~\cite{Mehta2021} for forced GL. 
This is likely the case of  vasomotion, where  neural stimulation acts as a drive for the oscillations of arteriole vessels~\cite{Drew2020,Broggini2024}.
Quantitative aspects of the neurovascular coupling remain mysterious, and will benefit from additional experiments. It would, in particular, be ideal to transiently shut down the neuronal activity and assess the undriven vasomotor dynamics. That will provide a sense of the elements that lead to locking effects, as recently reported in Ref.~\cite{Broggini2024}, and inform their modeling along the lines presented here.    
  
\medskip
\section{End Matter}

\paragraph{Numerical simulations.} 
We integrated Eq.~\eqref{eq:GL} by a finite-difference scheme with fourth-order discretization of the Laplacian. We simulated half of the spatial interval and imposed symmetry conditions ($\mathrm{Re}$/$\mathrm{Im}[A]$ even/odd) at $x=0$. We used $N=601$ collocation points, and the ratio of the time step $\Delta t$ to the smallest natural period satisfied $\frac{gL}{2\pi}\Delta t < 10^{-2}$. 
Data analysis was performed in the stationary regime eventually reached in simulations.

\paragraph{Clusters identification.} 

We identified plateaus by the clustering algorithm OPTICS~\cite{Optics2011}. Its input is the matrix $R_{ij} = 
       \frac{1}{T} 
       \sum_{t=0}^T
       \left|
            e^{i (\phi_i(t) - \phi_j(t))}
       \right|\,$ that measures the $T$-averaged coherence of the phases $\phi_{i,j}(t)$ at positions $i,j$ (see Fig.~S3 in \cite{NoteSM}). Three reasons motivated our choice of OPTICS. First, classification is not forced in uncertain cases and outliers, which 
       often correspond to defects in our case, are left unclassified. Second, it is density-based, which is handy for non-uniform extents of synchronization. Indeed, plateaus are not always ``clean'', i.e., with sharp edges and vanishing frequency differences along them~\cite{Akopov2003,NoteSM}, so that extended plateaus alternating with smaller ones and/or unlocked regions may be observed (see Fig.~S2 in \cite{NoteSM}). Third, the algorithm provides a systematic method to determine the number of clusters, which 
    was important to identify scaling~\eqref{eq:scalingN}.

\paragraph{Spectrum of the linearized GL equation.} 
Plugging the ansatz $A(\zeta,\tau)=\rho(\zeta)e^{\lambda \tau} $ in Eq.~\eqref{eq:BlochTorrey}, we obtain
\begin{equation}
\label{eq:eigenGL}
    (r^\prime -\lambda + i\alpha\zeta + \Delta)\rho = 0\,,
\end{equation}
with $\nabla \rho(\zeta = \pm 1/2) = 0$. Eq.~\eqref{eq:eigenGL} shows that its eigenvalues $\lambda_n$ are related to the $\sigma_n$'s of the Bloch-Torrey 
equation ($r'=0$) by the shift  $\lambda_n = \sigma_n + r^\prime$. We computed the $\sigma_n$'s by using a second-order discretization of the Laplacian in Eq.~\eqref{eq:BlochTorrey} and using the ZGEEV function in LAPACK \cite{CIT1999} to compute the eigenvalues. 
To span the $\alpha$ range, we fixed  $D=0.5$, $L=1$, and varied $g=10^{0.1n}$, with the integer $n\in [0,60]$. Our numerics is 
consistent with Ref.~\cite{Dyson1991} with the change that $L=2$ instead of our $L=1$. Results of \cite{Dyson1991} are re-derived in SM~\cite{NoteSM}.
 
\paragraph{Precursors of defects.} 
We consider the linearized GL Eq.~\eqref{eq:BlochTorrey} in the regime where the first branching occurred (see Fig.~\ref{fig3}a,b). We show that branching is associated with strong modulation of the amplitude (precursor of defects) around $x=0$. The pair of branched eigenvalues with the largest real part is denoted $\lambda_1^R\pm i \lambda_1^I$. The dominant asymptotics of the field $A$ at large times is
\begin{equation}
  A(x,t)\propto e^{\lambda_1^R\,t}\left[\rho_+(x)e^{i\lambda_1^I t}+\rho_-(x)e^{-i\lambda_1^I t}\right]\,,
  \label{eq:prodro}
  \end{equation}
where $\rho_{\pm}$ are the  eigenmodes. Eq.~\eqref{eq:eigenGL} for the eigenmodes is symmetric under $\lambda^I\mapsto -\lambda^I$, $x\mapsto -x$, $\rho\mapsto \rho^*$, which implies $\rho_-\left(x\right)=\rho_+^*\left(-x\right)$ \cite{Dyson1991}. The expression~\eqref{eq:prodro} at the origin reduces to $2e^{\lambda_1^R t} |\rho_+\left(0\right)| \cos\left[\lambda_1^I t+{\rm arg}\left(\rho_+\left(0\right)\right)\right]$, which vanishes periodically with period $2\pi/|\lambda_1^I|$. 

\medskip
\begin{acknowledgments}
We thank Jacob Duckworth, Vladimir Matchkov, and Marcelo Rozenberg for relevant discussions.  This work was supported by the NIH grants U19 NS123717 and 
U19 NS137920.

\end{acknowledgments}


%

\end{document}